\documentclass[onecolumn]{emulateapj}

\shorttitle{Relativistic solar protons and the GLE \#72}
\shortauthors{Augusto et al.}

\begin{document}


\title{Relativistic proton levels from region AR 12673 (GLE \#72) and the heliospheric current sheet as a Sun$-$Earth magnetic connection}


\author{C. R. A. Augusto, C. E. Navia\altaffilmark{1} and M. N. de Oliveira}
\affil{Instituto de Fisica, Universidade Federal Fluminense, 24210-346, Niteroi, Rio de Janeiro, Brazil}

\author{A. A. Nepomuceno}
\affil{Departamento de Ci\^{e}ncias a Natureza, Universidade Federal Fluminense,
28895-532, Rio das Ostras, RJ, Brazil}

\author{A. C. Fauth}
\affil{Instituto de F\'{i}sica Gleb Wathagin, Universidade Estadual de Campinas, 13083-859, Campinas, SP, Brazil}

\author{V. Kopenkin}
\affil{Research Institute for Science and Engineering, Waseda University, Shinjuku, Tokyo 169, Japan.}

\author{T. Sinzi}
\affil{Rikkyo University, Toshima-ku, Tokyo 171, Japan}


\altaffiltext{1}{E-mail address:navia@if.uff.br}


\begin{abstract}
On 2017 September 10 Neutron Monitors (NMs) apparatus located at ground level and high latitudes detected an increase in the counting rate associated to solar energetic particles (SEPs) emission from X8.2-class solar flare and its associated CME. This was the second-highest flare of the current solar cycle. The origin was the active region AR 12673 when it was located at the edge of the west solar disk, magnetically poorly connected with Earth. However, there was a peculiar condition: the solar protons accelerated by the CME shocks were injected within a
heliospheric current sheet (HCS) region when Earth was crossing this region. We show that often HCS and SEPs propagation are closely related. If the source locations of SEPs are within or close to HCS, the HCS play the role of a Sun$-$Earth magnetic connection. SEPs drift around HCS paths, and SEPs are also drift in a wide range of longitudes by the HCSs. In some cases, and especially when Earth crosses the HCS sector, a fraction of these particles can reach Earth with a harder energetic particle flux, triggering a ground-level enhancement (GLE). The blast on 2017 September 10, which triggered the GLE \#72, was the second in the current solar cycle. We show that the two GLEs, including all sub-GLEs observed in the current solar cycle, comes from solar explosions that happened within an HCS structure; this behavior is also observed in the GLEs of the previous solar cycle. In general, solar explosions from active regions poorly connected with Earth can trigger GLEs, through the
mechanism described above. In all cases, the SEPs drift processes by HCS structures provides an efficient particle transport, allowing the observation of these solar transient events.
\end{abstract}




\section{Introduction}
\label{intro}



Ground-level enhancements (GLEs), typically in the MeVGeV energy range, are sudden increases in cosmic ray intensities registered by neutron monitors (NMs), which are ground-based instruments that detect a variety of secondary particles, mainly neutrons produced by primary protons
penetrating the Earth’s atmosphere \citep{mpp08,gopa10}. These enhancements can be registered by other types of ground-based detectors, such as air shower detectors and muon telescopes \citep{l3c08,wang09,nitt12}. In most cases, GLEs occur during the intense X-class solar flares\footnote{Flares are classified into 5 major classes, X, M, C, B, and A, with X corresponding to the Geostationary Operations Environmental Satellite (GOES)  energy flux (in the energy band 100-800 pm) in excess of $10^4 \, Watts \,m^{-2}$ at Earth and successive classifications decrease over decades \citep{flet11}.} as well as the fast  (above $\sim 1000 \, km \, s^{-1}$) Coronal Mass Ejections (CMEs; \cite{gopa10,gopa12}). However, there are also some cases of GLEs associated with weaker flares and slower CMEs when preceded by other events that provides
seed particles that increase the efficiency of particle acceleration \citep{cliv06}. 

In general, the production of a GLE need harder energetic particle fluxes having strong fluences, above the background fluence due to galactic cosmic rays. Statistical analysis has shown that a conjunction between CME-driven interplanetary shock and flare are likely to be the cause of GLEs, but CMEs alone presumably does not cause GLEs \citep{firo10}.

On the other hand, a heliospheric current sheet (HCS) is a transition zone that separates regions of opposing interplanetary magnetic field polarity \citep{wilc65}. A minor geomagnetic activity is often observed when Earth crosses through a fold in the heliospheric current sheet, a vast undulating system of electrical currents shaped like the skirt of a ballerina. The Earth dips in and out of it all the time. These crossings are called ``solar sector boundary crossings'', and they occasionally trigger high-latitude auroras.

Galactic cosmic rays are also modulated by the HCS \citep{burg89,alan07}, particularly the effect of drift on the transport of cosmic rays due to an HCS that has been obtained by Jokipii \citep{joki81}. Drift effects are dependent on particle energy and on the Interplanetary Field (IMF) configuration. The deceleration of particles (protons) by drift processes on the $A>0$ configurations of the IMF retains a significant portion of their energies because, under this condition, the latitudinal drift is constraining. Galactic cosmic rays inferred from neutron monitors often show a small peak in the counting rate preceding the HCS crossing, followed by a drop after the crossing \citep{thom14}. Also, often an HCS structure has an influence on the propagation of a high-speed stream \citep{augu17}.

In addition, the effects of the HCS on the propagation properties of solar disturbances, such as the SEPs, depends on the source location in relation to HCS \citep{henn85,wei91,xie06}. The modulation of SEP also depends on the magnetic configuration ($A>0$ or $A<0$) of the IMF. SEPs near an HCS drift across a wider range of longitudes in the $A>0$ configuration, and drift across latitudes in the $A<0$ configuration \citep{batt18}.

In this paper, we show that the signals detected by spacecraft detectors on 2017 September 10 correspond to two different phases of the solar flare. The energy release, during the impulsive phase of the flare, was observed as an increase of gamma rays and hard X-rays registered by  the \textit{Reuven Ramaty High Energy Solar Spectroscopic Imager} (RHESSI) and \textit{Fermi} GBM. The GOES observed soft X-ray signals corresponding to gradual (or extended) phase of the flare. Solar protons observed by GOES  were accelerated by CME shock waves with an average speed of 948 km/s, triggering a radiation storm of up to S3 (strong) level in the NOAA storm scale. A conjunction between CME-driven interplanetary shocks and an X8.2-class flare triggered the GLE \#72 detected by neutron monitors located at high latitudes. Some characteristics of the GLE \#72 can be found in \citep{kurt18} and the kinematics of the blast on 2017 September 10 and the spectral characteristics of the associated energetic particles were reported in \citep{gopa18}.

However, as the blast happened in the  western edge of the solar disk, i. e., a non-geoeffective region because it is longitudinally separated by more than 90 degrees from Earth, the energetic particles (protons and ions) ejected needed to be drifted longitudinally in order to reach Earth. In the present case, this mechanism was enhanced, because protons were injected within an HCS, and under the $A>0$  configuration of the IMF (2017 September), besides particles drift across the HCS path,  they also drift in a wide range of longitudes. Thus, the HCS modulation has played the role of a Sun$-$Earth magnetic connection to obtain a harder energetic particle flux at Earth, triggering the GLE \#72.

These observations are reported, including a brief description of all  activity of sunspot AR 12673 in 2017 September, the origin of relativistic particle levels on 2017 September 10,  the connection between flare, CME, and SEPs, including the GLE \#72. In the Appendix, where we present five solar blasts that happened at the western edge of the solar limb, three of them the active region; at the time of the blast, it was no longer seen from Earth. Even so, in all cases, SEPs were observed, triggering GLEs or at least sub-GLEs. We also show two cases where the SEPs and HCS structures are correlated, even in events magnetically better connected to Earth. In all cases, the blasts occurred within an HCS structure. This means that SEPs propagation and HCS are closely related.



\section{Activity of Sunspot AR 12673 in 2017 September}
\label{activity}

\begin{figure}[th]
\vspace*{-1.0cm}
\hspace*{-0.0cm}
\centering
\includegraphics[clip,width=0.6
\textwidth,height=0.3\textheight,angle=0.] {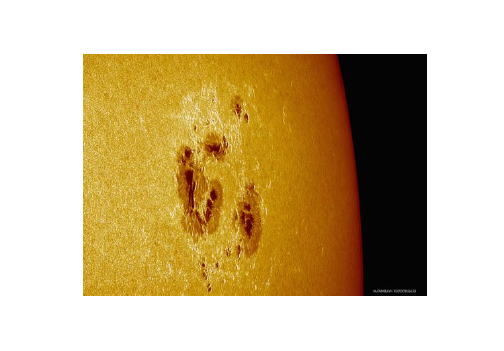}
\vspace*{-1.0cm}
\caption{Photograph of AR 12673,  after two days of the strongest eruption on 2017 September 6. It was in a ``beta-gamma-delta'' magnetic configuration. The AR 12673 region size was more than ten times the size of Earth. Reproduced with permission. \textcopyright Maximilian Teodorescu 2017.$^6$
}
\label{AR2673}
\end{figure}

The active region AR 12673 began to be visible on August 26, and during its rotation toward the sun's western limb, AR 12673 had a ``beta-gamma-delta'' magnetic configuration, producing strong eruption on the Sun. Indeed, AR 12673 erupted 25 M-Flares and 3 X-Flares, including the two largest of the current solar cycle (Cycle 24). Figure~\ref{AR2673} is a photograph of AR 12673 made by Maximilian Teodorescu of Magurele, Romania\footnote{\url{http://spaceweathergallery.com/indiv\_upload.php?upload\_id=139018}}, on 2017 September 8, i.e.,  two days after of the strongest eruption in the current solar cycle an X9.2-class flare, on 2017 September 6.

Two X-class solar flares erupted on 2017 September 6 from AR 12673. The first was a long-duration X2.2-class flare at 9:33 UT. The blast was associated with a narrow CME ejected on the western region of the solar disk. However, the average shock wave speed was only 419 km s$^{-1}$ and they were also injected mainly ``below'' of the ecliptic plane. This flare was the first X-class since 2015 May 5. The second was at 12:02 UT, and was also the strongest eruption in the current solar cycle, reaching the condition of X9.2-class flare. 

The blast was associated with a full-halo CME (CME \#0017, in the Cactus quicklook CME catalog), the lowest and highest shock velocities detected within the CME was 376 km s$^{-1}$ and  1955 km s$^{-1}$, respectively, with a median value of 978 km s$^{-1}$. The CME \#0017 was classified as halo IV because its angular width was above 270$^{\circ}$ and reached the value of 360$^{\circ}$. 

\begin{figure}[th]
\vspace*{-0.0cm}
\hspace*{-0.0cm}
\centering
\includegraphics[clip,width=0.8
\textwidth,height=0.4\textheight,angle=0.] {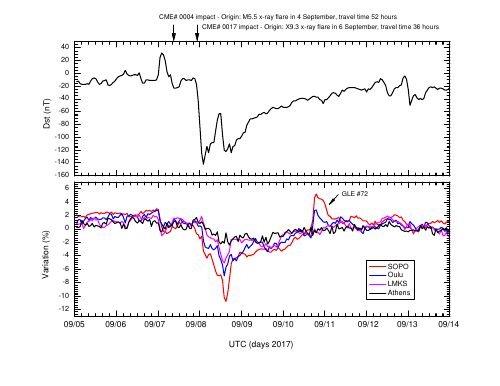}
\vspace*{-0.5cm}
\caption{Time profiles for nine consecutive days in 2017 September, in five different detectors. Top panel: the Dst geomagnetic index variation
Bottom panel: the counting rate (variation in percentage) in the South Pole, Oulu, Lomnicky (LMKS), and Athens NMs, respectively.}
\label{NM_9days}
\end{figure}

The presence of Type IV and Type II radio emissions associated with the blast means a strong coronal mass ejection was associated with solar radiation storms. After a travel of 35 h, the halo CME arrived at Earth on 2017 September 7 at $\sim$23:00 UT, triggering a strong G3 geomagnetic storm. This was the second strongest geomagnetic storm in the current solar cycle. The Dst geomagnetic index reached up to -142 nT in the transition from 2017 September 7-8; the top panel of Figure~\ref{NM_9days} summarizes the situation.

In addition, the strong geomagnetic storm triggered a Forbush decrease (FD). An FD is a transient decrease followed by a gradual recovery in the observed galactic cosmic ray intensity. The perturbed geomagnetic field during geomagnetic storms disperses the cosmic rays in the vicinity of the Earth, producing a fall in the counting rate at ground-level detectors. The FD intensity increases as the geomagnetic rigidity cutoff of the site of the detector decreases. Thus, the FDs are more intense in detectors located at high latitudes. The FD associated with G3 geomagnetic reached an intensity variation of up to  11\% at South Pole NM. The bottom panel of Figure~\ref{NM_9days} summarizes the situation.

On 2017 September 10, AR 12673 erupted when it was in the extreme western limb. It was the second strongest flare of the cycle 24, reaching up to X8.2-class, associated to a halo IV CME. A fraction of the shocks within the CME was ejected within an HCS region and when Earth was crossing the HCS region. The HCS allowed a direct magnetic (Sun$-$Earth) connection, because of the solar energetic particles, accelerated by the CME shocks  started arriving on Earth, up to reaching the S3 (strong) condition in the NOAA storm scale, triggering the second ground-level enhancement (GLE \#72) in the current solar cycle detected in some NMs, as shown in bottom panel in  Figure~\ref{NM_9days}. These observations are detailed in the next section.

\section{Origin of relativistic proton levels and the GLE \#72}
\label{origin}

\begin{figure}[th]
\vspace*{-3.0cm}
\hspace*{-0.0cm}
\centering
\includegraphics[clip,width=0.8
\textwidth,height=0.45\textheight,angle=0.] {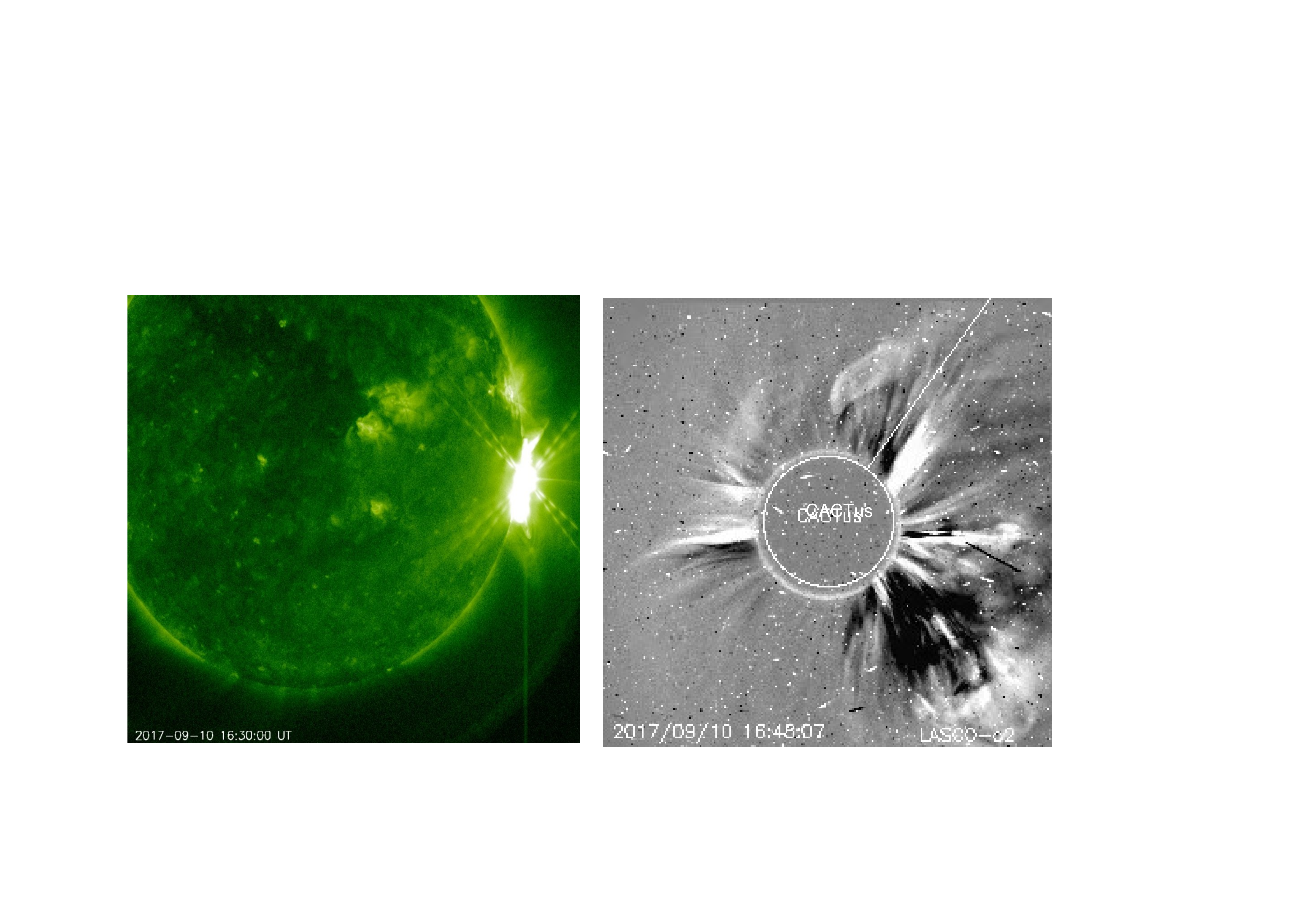}
\vspace*{-1.0cm}
\caption{Left panel: the extreme UV flash, of the second strongest solar flare in the current solar cycle, reaching X8.2-class, on September 10 at 16:06 UT from sunspot AR 12673 (Credit: Solar Dynamics Observatory). Right panel: The CME \#0037 eruption associated with the X8.2-class flare at 16:48 UT. It was a halo IV CME, reaching an angular width of 360$^{\circ}$ (full halo) at 16:48 UT. 
(Credit: Cactus analysis from LASCO detector).}
\label{flare}
\end{figure}

On 2017 September 10, almost out of view from our fair planet, rotating around the Sun's western edge, the active region AR 12673 erupted again at 15:35 UT, the blast was an X8.2-class flare.  An image in the extreme ultraviolet region was caught by Solar Dynamics Observatory and it is shown on the left panel of Fig.~\ref{flare}. The blast had its associated CME (CME \#0037 in the LASCO quicklook CME catalog)
and whose image captured by the Lasco on the SOHO spacecraft is shown on the right panel of  Fig.~\ref{flare}. This CME had a significant release of plasma and magnetic field from the solar corona, it was classified as halo IV and reaching the value of 360 degrees (full halo). 
The angular width derived from projected images is only an apparent quantity, it indicates the angular size of the CME volume projected onto the plane of the sky. A CME appears as a halo IV only if it is launched on a path close to the Sun$-$Earth direction and this happened only when the active solar region is near to the solar central region. But this is not the case for the CME \#0037, the active region that originated the blast was at the western edge. This effect it can be explained considering the reduction of the heliospheric pressure in the solar cycle 24 in relation to the previous (cycle 23), this can propitiate an anomalous expansion of coronal mass ejections near the Sun \citep{gopa15}. However, in this case, can be also associated, at least in part, to the location of the active region, within an HCS. Under this condition, the magnetic pressure of the environment region during the expansion of the CME is non-existent, because of the HCS is a transition region between two magnetic polarities, where $B\sim 0$.

\begin{figure}[th]
\vspace*{-0.0cm}
\hspace*{-0.0cm}
\centering
\includegraphics[clip,width=0.6
\textwidth,height=0.3\textheight,angle=0.] {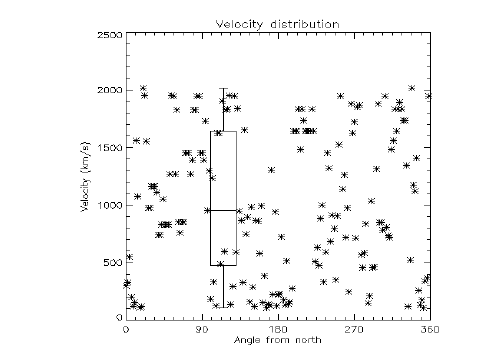}
\vspace*{-0.0cm}
\caption{CACtus (LASCO) distribution of the principal angle (latitude) of shock waves, measured counterclockwise from North (degrees) as a function of the shock waves velocity for the CME \#0037 (quicklook CME catalog) associated to the X8.2-class solar flare on 2017 September 10.
}
\label{cactus2}
\end{figure}

The data is from LASCO coronagraph images and automatically generated by CACTus \citep{robb09}. The $pa$ CACTus parameter correlated with the CME projected latitude, and it is defined as the middle angle of the CME when seen in the white-light images. The $pa$ parameter also depends on the orientation of the CME in the relation of the observer (Lagrange Point L1). The principal angle is measured counterclockwise from North (degrees). Thus, the projected latitudes are only a good estimation of the true direction of propagation. Values of the $pa$ close to 90 and 270 degrees represent zero latitudes. The $pa = 90^{\circ}$ and $pa = 270^{\circ}$ coincides with the eastern side and with the western side of the ecliptic plane, respectively \citep{augu18}. In addition, from Figure~\ref{cactus2} we can see that the shock wave velocities in the CME \#0037 ranging from 104 to 2013 km s$^{-1}$, with a median velocity of 948 km s$^{-1}$ and they have an almost uniform distribution, across the $pa$ region (this happens only for a full-halo CME).

\begin{figure}[th]
\vspace*{-0.0cm}
\hspace*{-0.0cm}
\centering
\includegraphics[clip,width=0.6
\textwidth,height=0.6\textheight,angle=0.] {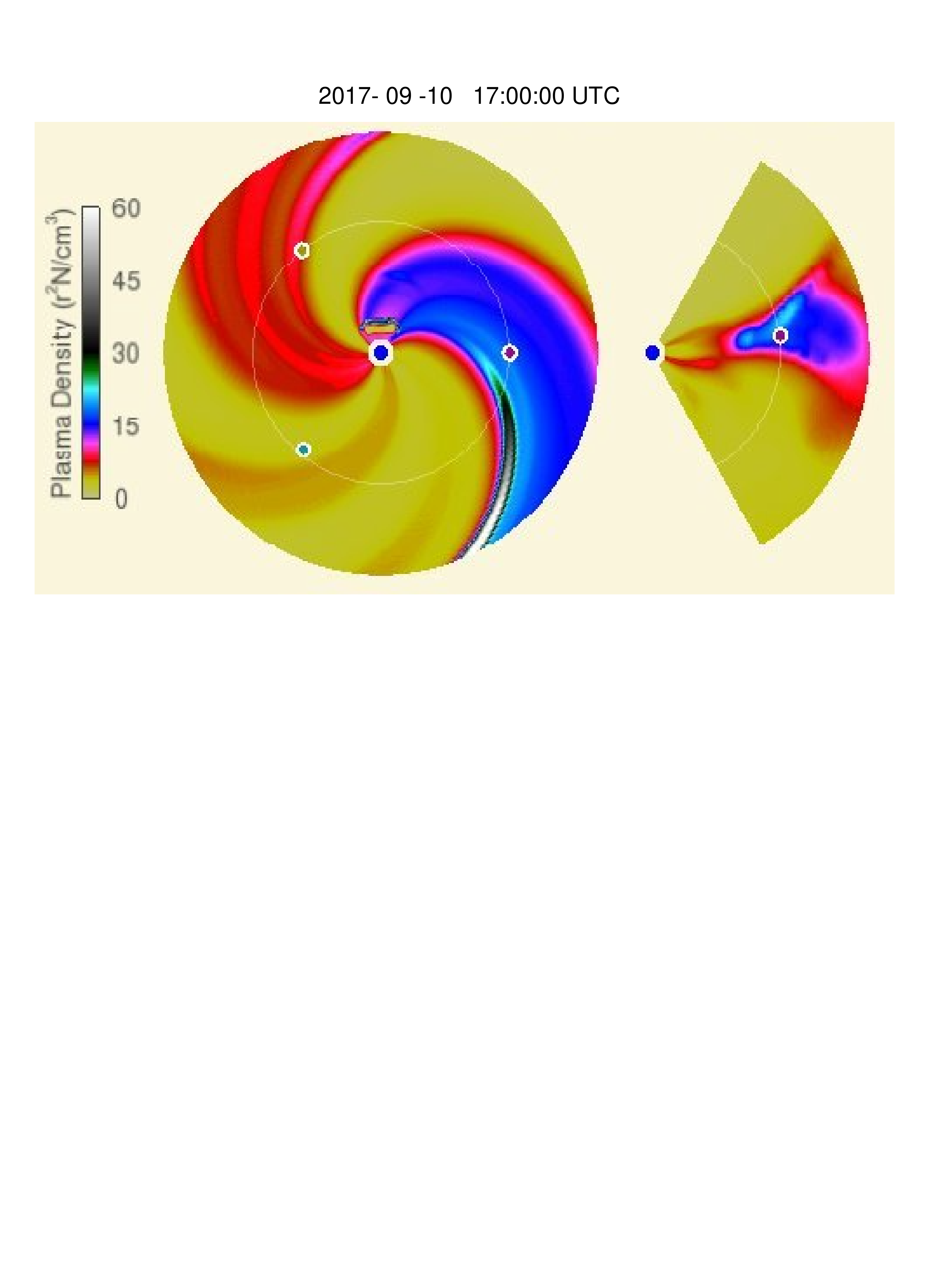}
\vspace*{-7.0cm}
\caption{WSA-Enlil model information on the density of solar wind for 2017 September 10, at 17:00 hours. Earth is marked as the red circle, the Sun is shown by the blue circle, and the green and yellow circles are the twin STEREO spacecraft in 1 AU orbits around the Sun. The circular image shows a view of the ecliptic plane viewed from above and the semi-circle a side view of this plane. Reproduced from NOAA. Image stated to be in the public domain.}
\label{wsa}
\end{figure}

\begin{figure}[th]
\vspace*{-0.0cm}
\hspace*{-0.0cm}
\centering
\includegraphics[clip,width=0.8
\textwidth,height=0.4\textheight,angle=0.] {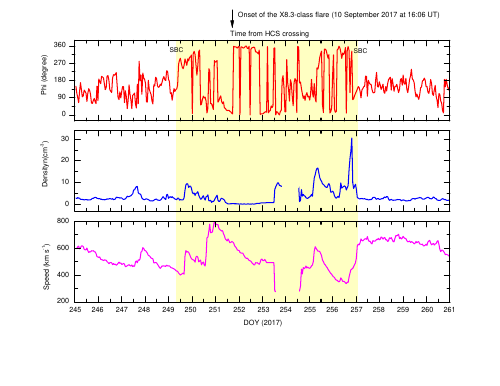}
\vspace*{-1.0cm}
\caption{Solar wind time profiles on 2017 September, from DOY 245 (September 2) to DOY 261 (September 18). From top to bottom panels: the $Phi$ angle, density, and speed. (Credit: ACE Solar wind data). The yellow area indicates the Earth-crossing time by the HCS, the sectors boundary crossing is indicated as SBC.}
\label{wind2}
\end{figure}

However, there was another important factor, a significant fraction of the CME was ejected within an HCS region. Figure~\ref{wsa}, shows a snapshot of the WSA-Enlil model\footnote{The WSA-Enlil model (\url{https:282//www.ngdc.noaa.gov/enlil\_data/2015/)}} run around on 10 September. The model shows the solar wind plasma density in the ecliptic plane. Blue colors show very high plasma density and grey color represents low density. The Earth is marked as the red circle, and the Sun is shown by the blue circle. Following this figure, we can see that on 10 September the Earth was crossing a wide region of high solar plasma density, i.e., crossing the HCS. The crossing of Earth, through the boundary sectors of HCS, is usually not associated with big disturbances in the geomagnetic field, but there is a signature on the solar wind parameters. In the present case, a sector boundary crossing (SBC) occurred in 2017 September 6 (DOY 249), when the magnetic field changed from ``away'' to ``towards'' Sun, and $\sim$ 8 days later (at transition from 13 to 14 September, DOY 256-257) back to ``away'', as shown in the time
profiles of Phi angle in the top panel of Figure~\ref{wind2}, there are also changes (fluctuations) in the solar wind density and speed, as shown in the central and bottom panel of Figure~\ref{wind2}, respectively. Data were taken from ACE. Particularly, Phi is the angle of the interplanetary magnetic field that is being carried out by the solar wind. Phi is measured in the GSM (geocentric solar magnetospheric) coordinate system. Phi would be 0 degree if it were pointing to the Sun and 180 degrees if it were pointing from the Sun to the Earth. Sudden and rapid changes in the Phi angle happen when the Earth is crossing an HCS sector. This property is used to limit the sector boundary crossing (SBC).

On the other hand, three dimensional Monte Carlo simulations on the propagation of solar energetic protons with energies from 1 to 800 MeV \citep{batt17,batt18}, has shown that the HCS has a big influence on the SEPs transport in the interplanetary magnetic field close to the ecliptic plane.

The results show that the presence of an HCS near high-energy particles play an important role in the longitudinal and latitudinal transport of SEPs, allowing it to drift over 180 degrees in longitude, i.e., protons can reach regions far from the injection longitude. Under the $A>0$ condition of the IMF, protons are drifted preferentially to eastern regions, whereas under the $A<0$ condition, protons are drifted 
preferentially to western regions.  However, in both cases, a fraction of protons are confined in the vicinity of the HCS. The study considered only the magnetic configuration at the solar minimum when the Sun polarity is well defined and consider only small injection regions.

However, there are at least two different conditions in the event of 2017 September 10, when compared with the simulated results. The first, is that the event happened during the declining phase of the current solar cycle, even so, close to the solar minimum. The second and more important is relative to the injection region of particles, in the event on 2017 September 10, particles were accelerated mainly by shock waves of a full-halo CME, due to an anomalous CME expansion, this means an extensive region when compared with the small injection region considered in the simulation.

Thus, if particles (protons) are injected within an HCS region and when the Earth is crossing by this region, the counting rate in the time profiles of detectors at Earth can be significantly increased. We believe that this mechanism happened on 2017 September 10, due to the X8.2-class flare and its associated CME ejected at the west solar edge and within an HCS, as are shown in Figure~\ref{flare} and Figure~\ref{wsa}.

Under a good magnetic connection and what happens when the active region which gave rise to the blast is at west, but not far from the Sun central region, solar energetic particles in the MeV to GeV energy region can bridge the Sun$-$Earth distance in a time as $\sim 10-20$ minutes. However, the delay between the blast beginning and the onset of proton flux increase observed by the GOES on 2017 September 10 was  $\sim 56$ minutes (see next section). A fraction of this delay time can be attributed to drift processes.

\section{Connection Flare CME and GLE}
\label{timing}

The observation by the Skylab around 40 years ago, of X-ray emissions by solar flares at the edge of the solar limb has enabled the clear identification up to three phases in a solar flares \citep{pall77}, a precursor phase, an impulsive phase (prompt) and a gradual phase (delayed). However in most cases, only the two last phases are identified, the impulsive and the gradual according to the their X-ray duration 
\citep{ruff97}. From the analysis of the timing of the diverse populations in large events, is possible to identify these two types, in the same explosion. They are known as different phases of a solar flare \citep{lin02,lin03}.

In some flares, the timing of electromagnetic emissions and relativistic protons suggests that the first proton peak is related to the acceleration during the impulsive phase \citep{murp87,klei14}. Protons and heavier ions are accelerated in the impulsive phase to relativistic energies by small-scale coronal loops.  The impulsive phase is characterized by a fast rise and shorter decay times, with duration of some tens of minutes \citep{pall77}. A fraction of these high-energy particles can interact with the nuclei of the different elements in the ambient solar atmosphere, these interactions can produce neutral pions which decay immediately and generate a broad gamma-ray line with a maximum near 70 MeV \citep{kurt13}. There is also the emission of hard X-rays in this phase, due to bremsstrahlung, produced by electrons
that have been accelerated to much higher energies than those found in the ambient plasma \citep{lin03}. The hard X-rays and the gamma-rays emissions are the evidence of acceleration of electrons and protons (ions) to relativistic energies in this impulsive phase.

\begin{figure}[th]
\vspace*{-0.0cm}
\hspace*{-0.0cm}
\centering
\includegraphics[clip,width=0.8
\textwidth,height=0.6\textheight,angle=0.] {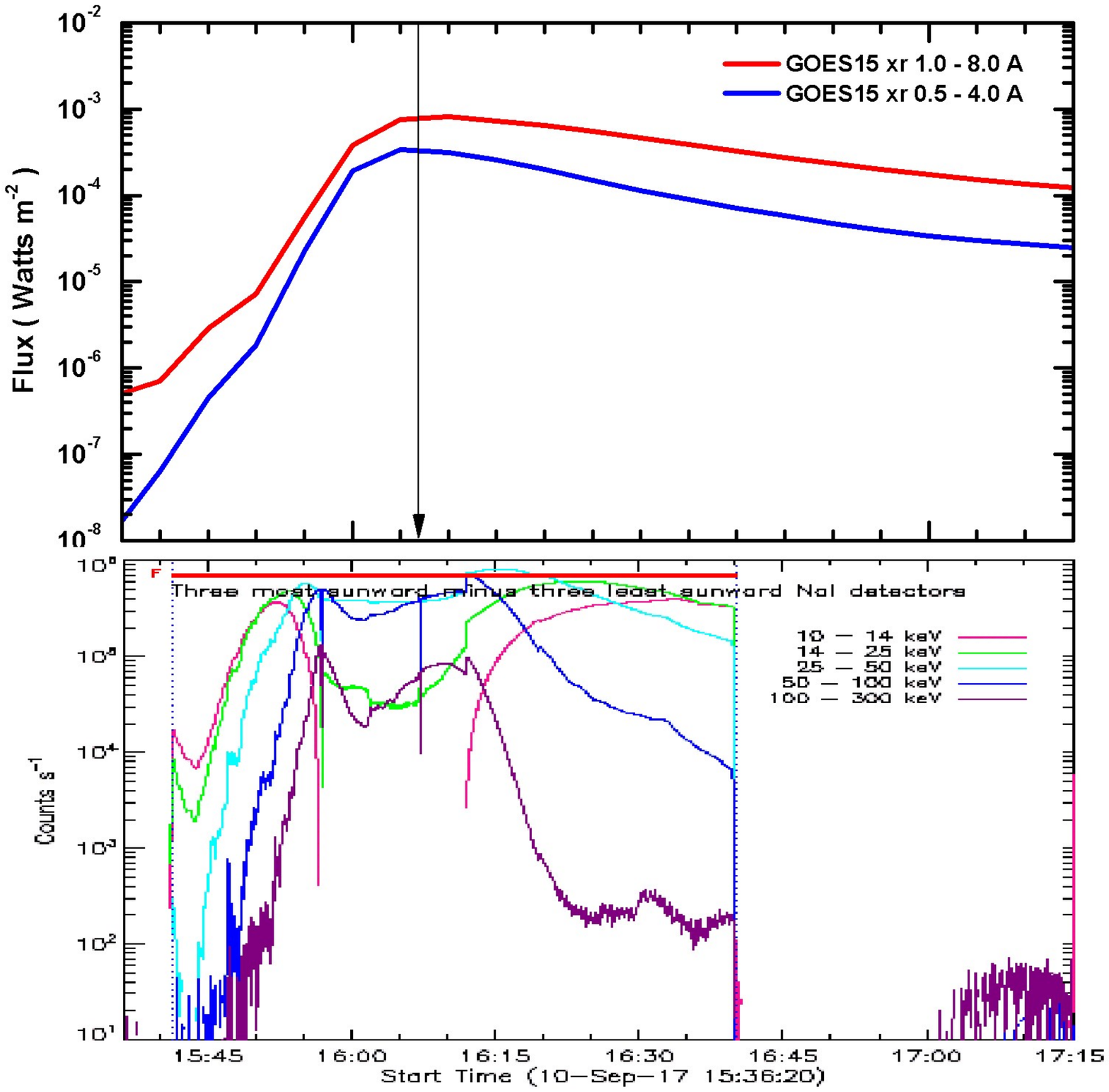}
\vspace*{-2.0cm}
\caption{Time profiles on 2017 September 10, in two different detectors. Top panel: the GOES 15 soft x-ray flux at two wavelengths. Bottom panel: the Fermi GBM hard X-ray quicklook on \textit{RHESSI} orbit times.}
\label{fermi}
\end{figure} 

In general, the peak of the impulsive phase coincides with the rise time of the soft X-rays due to the CME expansion and indicates the gradual phase of the solar flare. Protons and ions are accelerated at this gradual phase by CME shocks. Thus, these particles (SEPs) are spread over a broad region in solar longitude and under some (geoeffective) conditions of the active area, or if it to be inside or close to an HCS structure, the most energetic particles can give rise to a GLE \citep{chup09}.

The relativistic particles emission in the large particle event on 2017 September 10, is also likely to be interpreted in two different phases, an impulsive (prompt) and a gradual (delayed) particle acceleration.

Signatures of the impulsive phase were seen by \textit{RHESSI} \citep{lin02} in hard X-ray and gamma-ray emissions up to 20 MeV. The onset time of the \textit{RHESSI} signal is estimated by  as 15:50 UT, with a peak at 15:57 UT\footnote{\url{http://sprg.ssl.berkeley.edu/\~tohban/browser/?show=grth+qlpcr+gbmo+gbmd}}.

A comparison between the hard X-ray emission during the impulsive phase as observed by Fermi GBM and the soft X-ray emission during the gradual phase as observed by GOES 15 is shown in Figure~\ref{fermi}. From this figure, we can see that the hard X-ray emission, peaked about 10 minutes before then the soft X-ray peak.

\begin{figure}[th]
\vspace*{-1.0cm}
\hspace*{-0.0cm}
\centering
\includegraphics[clip,width=0.8
\textwidth,height=0.6\textheight,angle=0.] {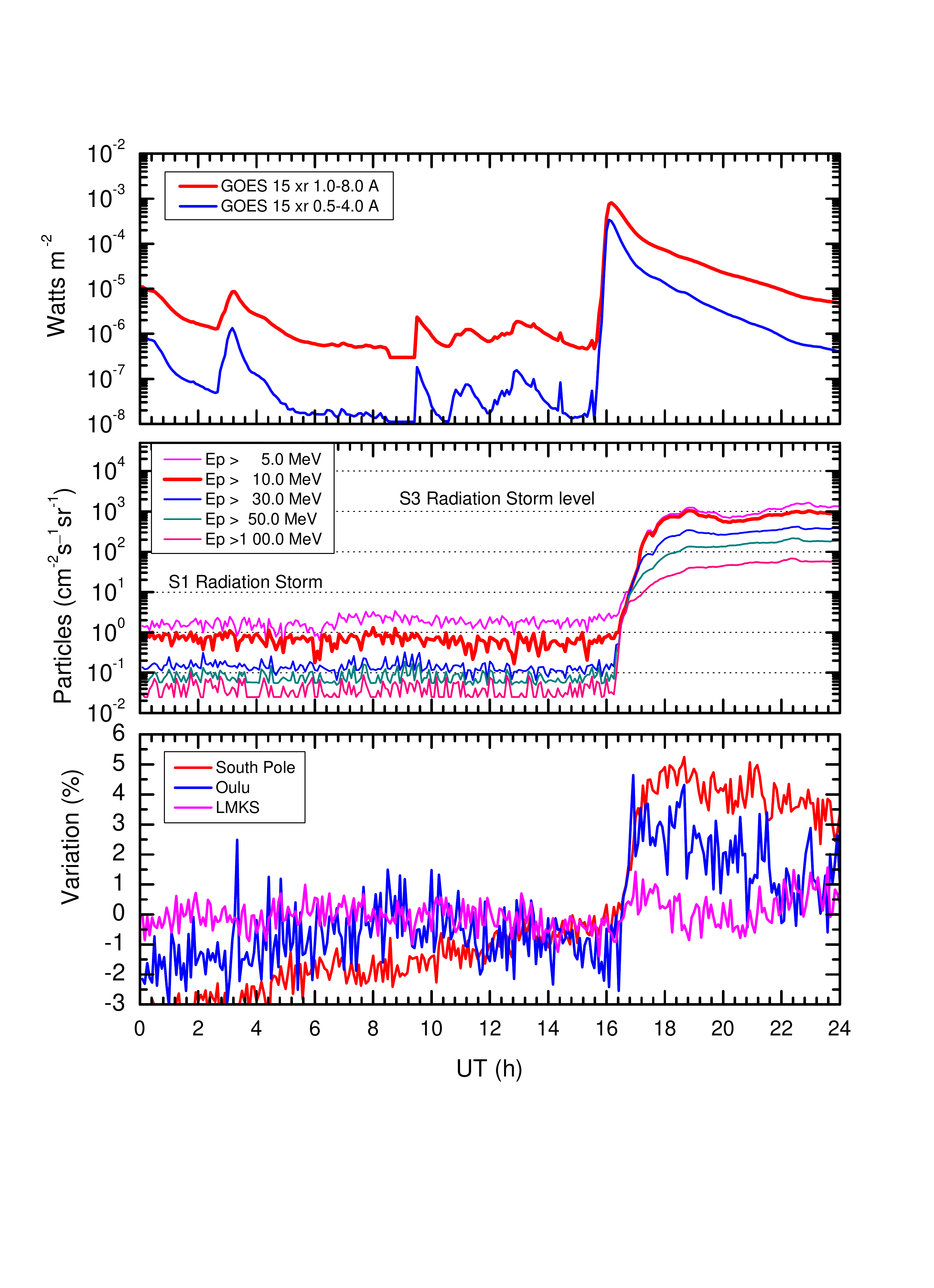}
\vspace*{-1.5cm}
\caption{Time profiles on 2017 September 10, in four different detectors. Top panel: the GOES 15 x-ray flux at two wavelengths. Central panel: the GOES proton flux in five energy band. Bottom panel: the counting rate (variation in percentage)in South Pole, Oulu, and Lomnicky (LMKS) NMs, respectively.}
\label{flare_xr}
\end{figure}

In most cases, a type II and IV radio emissions are observed in the gradual phase at the high corona \citep{gopa07} and they are typically associated with strong coronal mass ejections and solar radiation storms. In fact, in the event around Sunspot AR 12673 on 10 September, there were type II radio emissions, with onset at 16:08 UT, indicating that the coronal mass ejection was associated with the X8.2-class observed in the soft X-ray emission by GOES 13, as shown in  Figure~\ref{flare_xr} (top panel).

The blast generated a strong S3 Level Solar Radiation Storm (a proton flux at 1AU above 1000 particles per cm$^2$ per second and energies above 10 MeV), as observed by the GOES satellite. Figure~\ref{flare_xr} (central panel) summarizes the situation, showing the GOES proton flux for five different energy band, from 5 MeV to 100 MeV. We can see a fast rise increase in relativistic proton levels and after 3 hours of the onset (16:10 UT), the solar radiation storm reaches the S3 level (strong level) in the NOAA storm radiation scale.

Correlated with this observations we have the signature of the GLE \#72 (Fig.~\ref{flare_xr} (bottom panel)), in the time profiles in the counting rate of three neutron monitors: South Pole (0.1 GV)(confidence 5.6\%), Oulu (0.8 GV)(confidence 3.5\%, and Lomnicky (LMKS) (3.8GV) (confidence 0.7\%). As expected, neutron monitors in places with high magnetic rigidity cutoff, for instance, the Athens NM (8.5 GV) did not detect any signal (see Figure~\ref{NM_9days}). 


\section{Conclusions}
\label{conclusion}

On 2017 September 10, an X8.2-class solar flare erupted from the active region AR 12673. This was the second strongest solar flare of the current solar cycle. Besides, the blast happened under three circumstances, the first was the location of the AR 12673, on the western edge of the solar disk, without a magnetic connection with the Earth. The second is the AR 12673 footprint, within an HCS region and the third was a temporal coincidence, the blast happened when the Earth was crossing by this HCS region. 

The 2017 September 10 event, here analyzed, reinforce the hypothesis that SEPs propagation is closely related with the HCS structures. Active regions at the solar limb have a poor magnetic connection with Earth, but, if the blasts are within an HCS structure, SEPs from these blasts, can reach Earth and triggered GLEs.

This means that the HCS structure plays the role of a Sun$-$Earth magnetic connection. We present in the Appendix four GLEs (2001 April 18, 2014 January 6, 2012 May 17 and 2014 November 1) whose origins were at the western edge of solar limbo and in all cases, the blasts happened within an HCS structure and when the Earth was crossing an HCS structure (except the event on 2014 January 6).

The same mechanism can be seen in the event on 29 October 2015 \citep{augu16}, SEPs triggered a sub-GLE when the Earth was crossing through a fold of the heliospheric current sheet. In short, the two GLEs and all sub-GLEs, so far, that happen in the current solar cycle have this behavior, i.e., an HCS structure as a Sun$-$Earth magnetic connection.

Also, besides the event on 2001 April 18, that happened in the previous solar cycle, we show also two GLEs \#68 and \#69 in January 2005, they were detected when the Earth was crossing by an HCS structure. Both were originated from the same active region when it was within an HCS structure at the western region of the solar disk, but not at the edge of the solar disk. This means that even the SEPs from solar active areas, better connected to Earth, the HCS structures play a role of an efficient magnetic connection.

From a timing analysis, we found that the signals detected from this blast on 2017 September 10, correspond to different phases of the solar flare. The energy release, during the impulsive phase of the flare, was observed as an increase of gamma rays and hard X-rays, registered by \textit{RHESSI} and Fermi GBM and peaking at $\sim$ 15:57 UT.  While the energy release in the gradual phase of the flare was observed as an increase of  soft X-rays registered by  GOES and peaking at 16:06 UT. The GOES proton flux also corresponds to energy release in the gradual phase of the flare. Solar protons were accelerates by shock waves with a median speed of 948 km s$^{-1}$, reaching a radiation storm of up to S3 (strong) level. In addition, the ground-based detectors located in high latitudes observed a GLE (the GLE\#72) with a confidence of up to 5.6\%, in temporal correlation with the GOES proton flux. 

Finally, some peculiarities of the active region AR 12673 were reported, such as the eruption on 6 September, an X9.3-class flare, the strongest flare of the current cycle. It was associated with a halo CME toward Earth, triggering the second major geomagnetic storms of the current solar cycle (so far) on 7 September. 

\section*{Appendix A.}

In this appendix, we show seven solar flares associated with CMEs that triggered GLEs, five of them happened on the western edge of the solar disk, and in three of them, the source of the events, i.e., the active region, already had turned to another side of the Sun and it was no longer visible from Earth. 

We highlight that in all cases, the solar active regions at the time of the explosions, were within an HCS structure and in six cases, the blast happened when the Earth was crossing this region. These events have very similar characteristics to the event on 2017 September 10 here analyzed. 

These flares and their associated CMEs produced SEPs with harder energetic particle flux, triggering five GLEs (four of them in the previous solar cycle) and two sub-GLEs. Thus, the structure HCS is able to favor an efficient transport of these relativistic particles to the Earth.

\begin{figure}[th]
\vspace*{-0.0cm}
\hspace*{-2.0cm}
\centering
\includegraphics[clip,width=1.1
\textwidth,height=0.7\textheight,angle=0.] {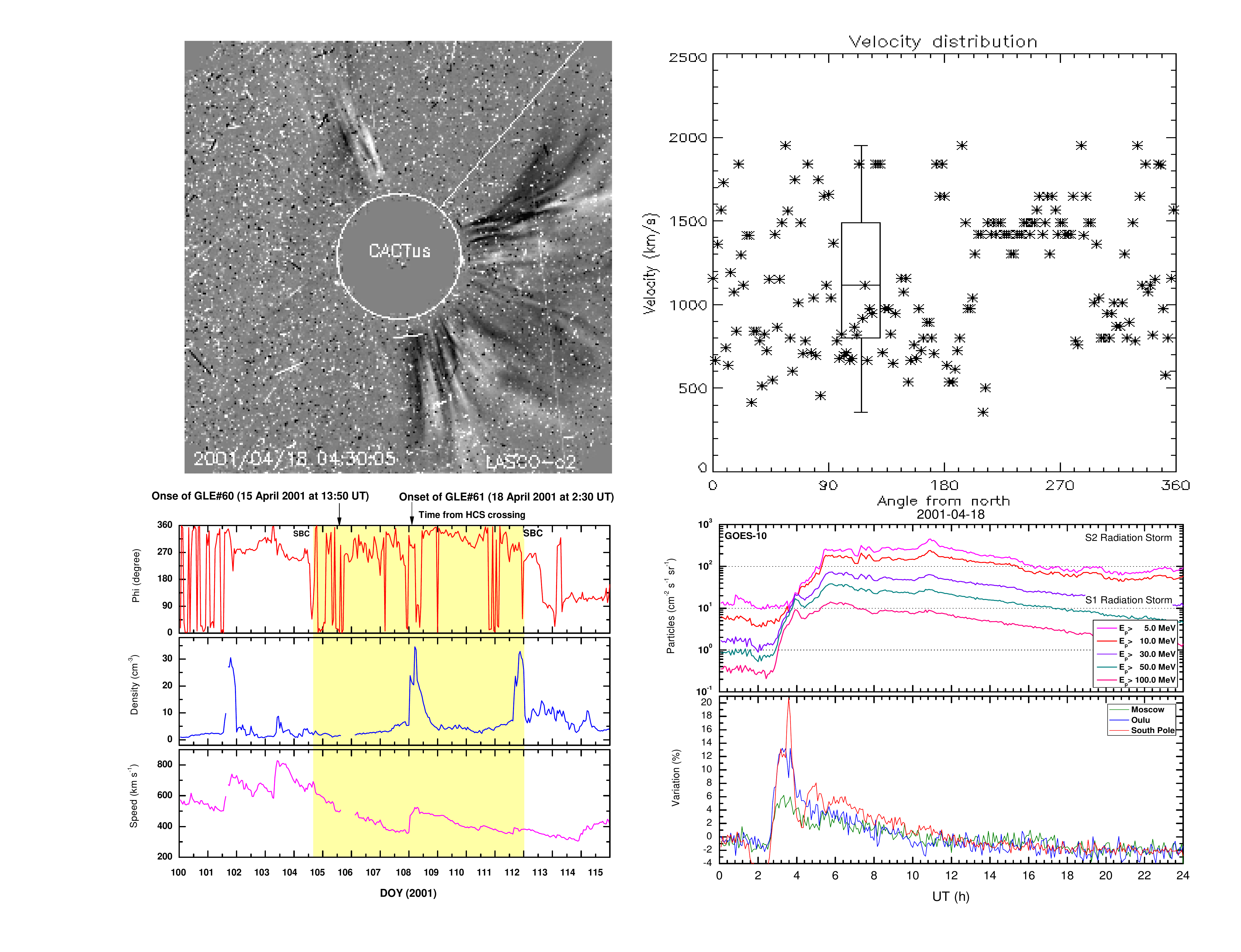}
\vspace*{-0.0cm}
\caption{Left top panel: the CME \#0098 eruption on 2001 April 18. It was a halo IV CME, reaching an angular width of 360 degrees (full halo) at 02:54 UT. (Credit: Cactus analysis from LASCO detector) Right top panel: CACTus (LASCO) distribution of the principal angle (latitude) of shock waves, measured counterclockwise from North (degrees) as a function of the shock waves velocity for the CME \#0098. Left bottom panel: Solar wind time profiles, the $Phi$ angle, density, and speed, respectively. The vertical arrows indicate the onset of GLE \#60 and GLE \#61, on 2001 April 15$-$18. The yellow area indicates the Earth-crossing time by the HCS, the sectors boundary crossing is indicated as SBC. Right bottom panel: the proton GOES flux for six energy bands and the counting rate of GLE \#61 in several neutron monitors.}
\label{fig9w}
\end{figure} 

\subsection*{A.1 Events on 2001 April 15 and 18}

The most spectacular solar blast from active region AR 19415 was on 2001 April 18 because, despite to be located behind the solar western limb at 02:54 UT, emerged a full-halo CME.

The CME  is the 0098 in the CACTus catalog, with shock wave velocities from 359 to 1953 km s$^{-1}$ and a median value of 1116 km s$^{-1}$, as shown in Figure~\ref{fig9w} (left top and right top panel). This behavior is hard to understand. How an occulted region can generate a full
halo CME. In addition, it also difficult to understand how a poorly connected region can be the origin of SEPs triggering a GLE, the GLE \#61.


In 2001, the WSA-Enlil model of the heliosphere, that predict the solar wind structure was not available yet. However, from the solar wind parameters obtained from ACE spacecraft at L1, is possible to see that at time blast, on 2001 April 18 (DOY 108), the Earth was crossing a wide HCS region (yellow region, in Figure~\ref{fig9w} (left bottom panel)). 

Because the HCS region was very wide, the active region also was within the HCS structure. This means, that the CME expansion was in a region where the environment magnetic field is close to zero, this propitiated an anomalous expansion of coronal mass ejections, it was a full halo CME as seen from Earth. In addition, the SEPs from the blast were drifted around the HCS path, reaching Earth and triggering the GLE \#61, as shown in Figure~\ref{fig9w} (right bottom panel).

We would like to point out, that three days before, on 15 April 2001 (DOY 105), the AR 19415 already had blasted SEPs from this explosion triggered the GLE \#60. This happened when AR 19415 already was within the HCS region (see Figure~\ref{fig9w} (left bottom panel)).

 \begin{figure}[th]
\vspace*{-0.0cm}
\hspace*{-1.0cm}
\centering
\includegraphics[clip,width=1.1
\textwidth,height=0.6\textheight,angle=0.] {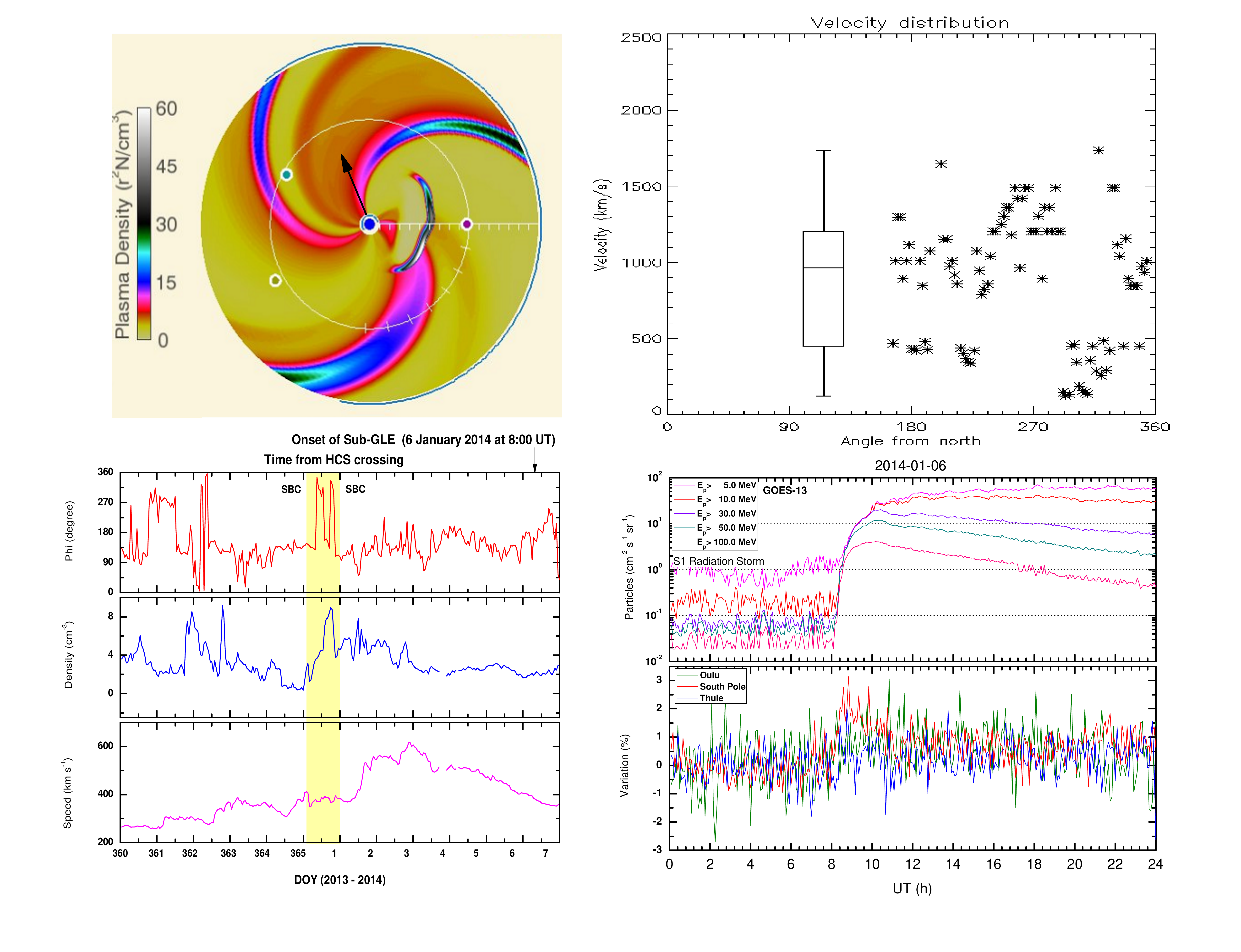}
\vspace*{-0.0cm}
\caption{Left top panel: WSA-Enlil model information on the plasma density of solar wind for 6 January 2014, the black arrow indicates the flare direction. Right top panel: CACtus (LASCO) distribution of the principal angle (latitude) of shock waves, measured counterclockwise from North (degrees) as a function of the shock waves velocity for the CME \#0028 associated to the solar flare on 6 January 2014.
Left bottom panel: Solar wind time profiles, the $Phi$ angle, density, and speed, respectively. The vertical arrow indicates the onset of sub-GLE, on 6 January 2014. The yellow area indicates the Earth-crossing time by the HCS, the sectors boundary crossing is indicated as SBC. Right bottom panel: The proton GOES flux for five energy bands and the counting rate of the sub-GLE in several neutron monitors.
}
\label{fig10w}
\end{figure} 

\section*{Appendix B. \\ Event on 2014 January 6}

On 2014 January 6 at aproximately 07:50 UT the GOES satellite detected a soft X-ray emission, was a C2-class flare from a solar eruptions located behind western solar limb, i.e., the source of the flare, the active region AR 11936 could no longer be seen from Earth. The tip from flare were also seen by SDO/AIA as a filament. For this flare the STEREO observations indicates that the flare would be a X3.5-class at 08:00 UT.


The flare had its associated CME, registered by LASCO as the CME 0028 in the CACTus catalog, with a width of 188 degrees and shock waves with velocities from 123 to 1736 km s$^{-1}$ with a median value of 961 km s$^{-1}$, as shown in the Figure~\ref{fig10w} (right top panel).

As in the previous case, the source (AR 11936) at time of the blast was within an HCS structure, as shown in Figure~\ref{fig10w} (left top panel) where the predicted plasma density of solar wind structure is show according to WSA-Enlil model, the black arrow indicate the CME direction emission.

However, at the time of the blast the Earth already had crossing the HCS, see Figure~\ref{fig10w} (left bottom panel), where the time series of solar wind parameters is presented. Even so, the HCS played the role of a Sun$-$Earth magnetic connection, because the SEPs were drifted in a wide region of latitudes, following the HCS structure. Thus SEPs began to reach the Earth at 08:00 UT, triggering an S1-class solar radiation storm, as shown in Figure~\ref{fig10w} (right bottom panel). The SEPs triggered a sub-GLE detected by several neutron monitors, those with a low geomagnetic rigidity cutoff, i.e., located at high latitude (polar regions), as shown in Figure~\ref{fig10w} (right bottom panel). 
 
An analysis of this event is presented in \citep{thak14}.

\begin{figure}[th]
\vspace*{-0.0cm}
\hspace*{-1.0cm}
\centering
\includegraphics[clip,width=1.1
\textwidth,height=0.6\textheight,angle=0.] {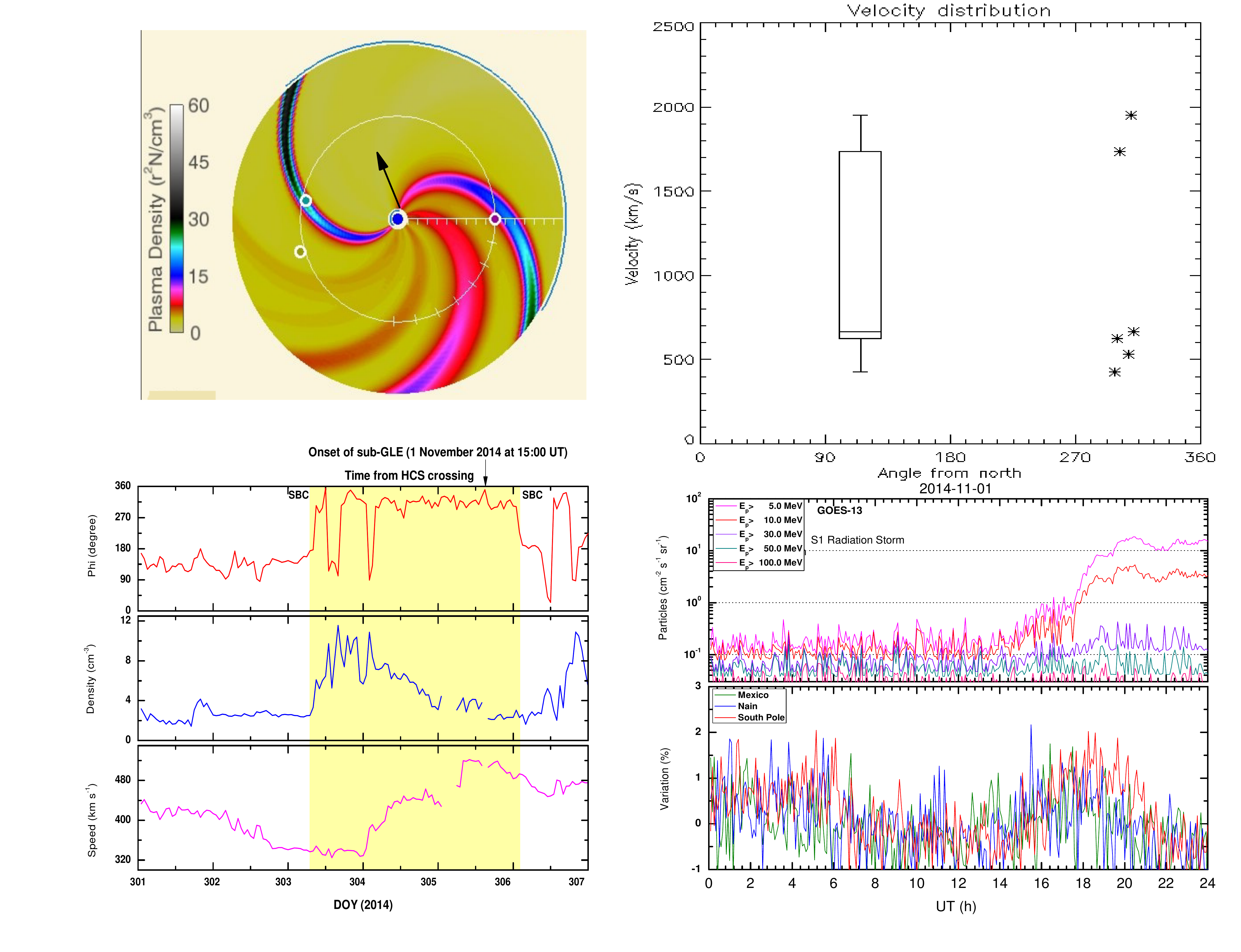}
\vspace*{-0.0cm}
\caption{Left top panel: WSA-Enlil model information on the plasma density of solar wind for 2014 November 1, the black arrow indicates the flare direction. Right top panel: CACtus (LASCO) distribution of the principal angle (latitude) of shock waves, measured counterclockwise from North (degrees) as a function of the shock waves velocity for the CME \#0003 associated to the C4.5-class solar flare. Left bottom panel: Solar wind time profiles, the $Phi$ angle, density, and speed, respectively. The vertical arrow indicates the onset of sub-GLE, on 2014 November 1. The yellow area indicates the Earth-crossing time by the HCS, the sectors boundary crossing is indicated as SBC. Right bottom panel: the proton GOES flux for six energy bands and the counting rate shown the sub-GLE in several neutron monitors.
}
\label{fig11w}
\end{figure}

\subsection*{B.1. Event on 2014 November 1}

The solar active region AR 12192 was the biggest sunspot, so far, of the current solar cycle, and on 2014 November 1 it was no longer visible from Earth. However, even under this condition, a soft X-ray emission was detected by GOES satellite from AR 12192 at 9:20 UT, probably was an X-class flare, while from Earth was observed only as a C4.5-class.

Another important factor is shown by the images obtained through the  Lasco-c2 equipment. The CACTus catalog reveals that the CME 0003 (associated with this blast) was a narrow CME, with an angular width of only 14 degrees, even so,  with shock waves on the western side of the ecliptic plane, as show in the Figure~\ref{fig11w} (right panel), we can see that the velocity of the shocks produced during the CME ejection reached  $\sim 2000$ km s$^{-1}$.

The combination of the shock waves capable of accelerating particles in the range of up to GeV, with the structure HCS that is able to favor the
transport of these relativistic particles to the neighboring of the Earth. Showing that AR 12192 eruption is the source of the SEPs, reaching near an S1-level radiation storm on November 1, as shown in Figure~\ref{fig11w} (left bottom panel). The SEPs have triggered a sub-GLE observed by neutron monitors with a low geomagnetic rigidity cutoff, i.e., those located at high latitudes (polar regions), as shown in Figure~\ref{fig11w} (right bottom panel).

\subsection*{B.2. Event on 2012 May 17}

On 2012 May 17 at 01:25 UT, an M5.1-class flare exploded in the edge of the western solar limb. A CME was also associated with this flare. About 20 minutes after the soft X-ray emission, solar particles reached the Earth, there was a proton flux with energies above 100 MeV, leading to an S2 solar radiation storm level. 

This explosion was the source of the first GLE  (GLE \#71) since 2006 December and the first of the current solar cycle 24. The GLE was detected by neutron monitors (NM) and other ground-based detectors. Analysis of the GLE \#71 on 2012 May 17 using data from the global neutron monitor network is presented in \cite{mish14,gopa13}.

\begin{figure}[th]
\vspace*{-0.0cm}
\hspace*{-0.0cm}
\centering
\includegraphics[clip,width=0.9
\textwidth,height=0.5\textheight,angle=0.] {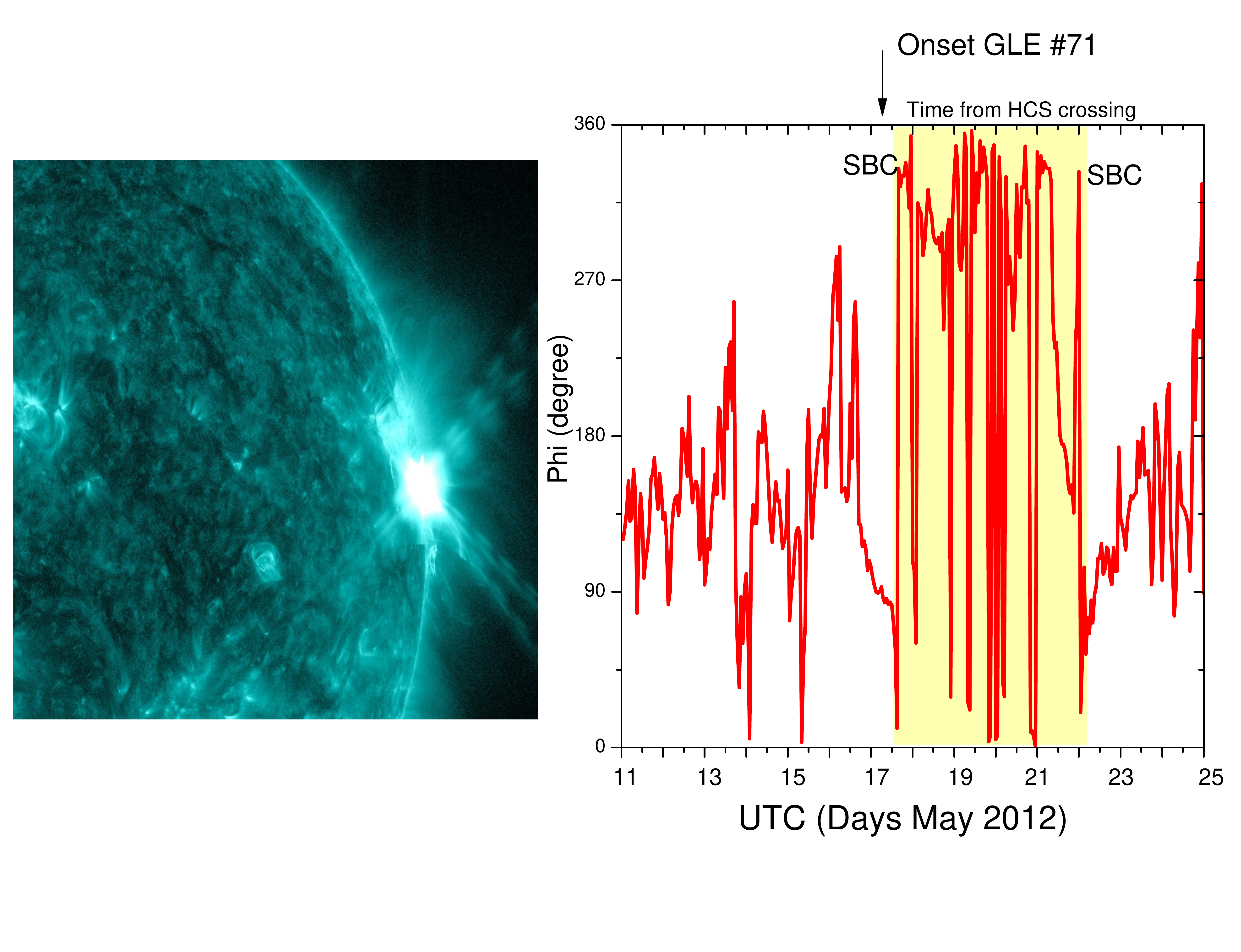}
\vspace*{-1.0cm}
\caption{Left panel: the extreme UV flash, of the solar flare reaching M5.1-class, on 2012 May 17 at 01:25 UT (Credit: Solar Dynamics Observatory). Right panel: the phi angle time profiles for seven consecutive days. The yellow area indicates the Earth-crossing time by the HCS, the sectors boundary crossing is indicated as SBC. The vertical arrow indicate the onset time of GLE \#71.}

\label{fig12w}
\end{figure} 

We would like to highlight that this event also reinforces the hypothesis that SEPs and HCS structures are closely related. The explosion (a medium sized one) from an active region even with a poor connection with the Earth triggered a GLE. We believe that this occurrence was possible due to the presence of an HCS structure, as shown in Figure~\ref{fig12w} (right panel), where the time profiles of phi angle is shown, there is a clear demarcation of time when Earth is crossing by the HCS structure (yellow area). In short, the HCS plays the role of a magnetic connection.

\begin{figure}[th]
\vspace*{-0.0cm}
\hspace*{-0.0cm}
\centering
\includegraphics[clip,width=0.8
\textwidth,height=0.4\textheight,angle=0.] {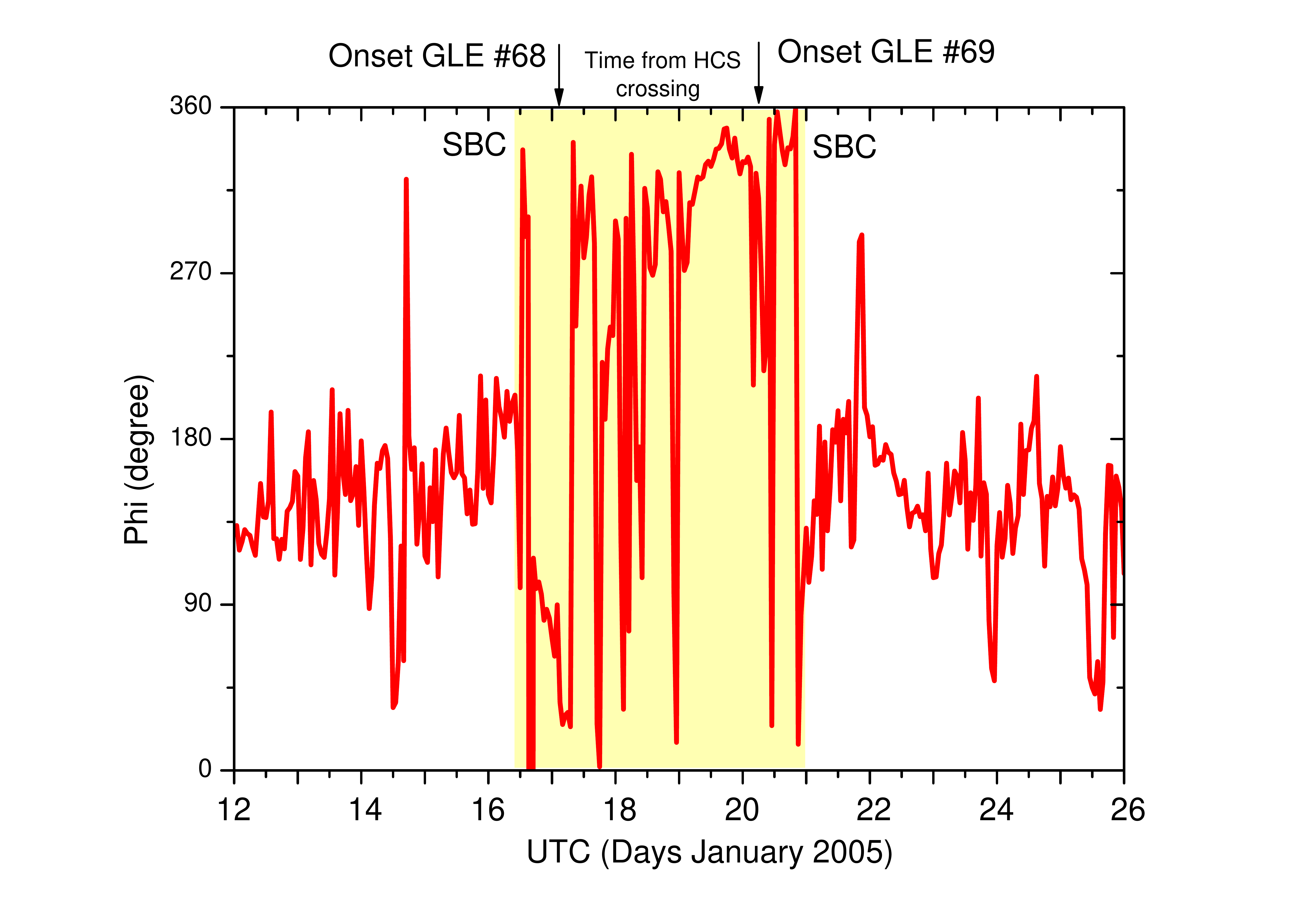}
\vspace*{-0.5cm}
\caption{Phi angle time profiles for seven consecutive days. The yellow area indicates the Earth-crossing time by the HCS, the sectors boundary crossing is indicated as SBC. The vertical arrows indicate the onset time of GLE \#68 and GLE \#69, respectively.
}
\label{fig13w}
\end{figure} 

\subsection*{B.3. Events on 2005 January}

On 2005 January 17, a giant sunspot 720 has unleashed another big solar flare. The X3-class explosion hurled a CME directed to the Earth. Protons accelerated up to relativistic energies by the blast, hit the Earth and triggering the GLE \#68. 

After around three days, the giant sunspot 720 erupted again on 2005 January 20, unleashing a powerful X7-class solar flare. The blast hurled a coronal mass ejection (CME) into space and sparked the strongest radiation storm since October 1989, triggering the GLE \#69.

We would like to highlight again, that these two events also reinforce the hypothesis that SEPs and HCS structures are closely related. Despite in both cases, the active region that generated these two explosions had a better connection with the Earth, also in both cases,
the GLEs were detected when the Earth was crossing by the HCS sector. Due to the large width of the HCS sector, the two explosions happened within the HCS structure,  as shown in Figure~\ref{fig13w}, where the time profile of phi angle is shown, as in the previous case, there is a clear demarcation of time when Earth is crossing by the HCS structure (yellow area). 

\section{ACKNOWLEDGMENTS}

We acknowledge the use of the NMDB database, founded under the European Unions FP7 programme (contract no. 213007). The data of the South Pole NM is provided by the University of Delaware with support from the U.S. National Science Foundation under grant ANT-0838839. We express our gratitude to the ACE Science Center, the NOAA Space Weather Prediction Center for valuable information and data used in this study , and Maximilian Teodorescu for the AR 12673 image. This work is supported by the National Council for Research (CNPq) of Brazil, under Grant 306605/2009-0 and 308494/2015-6, Rio de Janeiro Research Foundation (FAPERJ), under Grants 08458.009577/2011-81 , and So Paulo Research Foundation (FAPESP), under Grant 2011/50193-4.
 
\newpage
\newpage
\vspace*{+1.0cm}

\end{document}